\documentclass[%
 reprint,
 amsmath,amssymb,
 aps,
]{revtex4-2}

\usepackage{graphicx}
\usepackage{dcolumn}
\usepackage{bm}
\usepackage{array}
\usepackage{booktabs}
\usepackage{makecell}
\usepackage{color}

\begin{document}
\newcommand{\PreserveBackslash}[1]{\let\temp=\\#1\let\\=\temp}
\newcolumntype{C}[1]{>{\PreserveBackslash\centering}p{#1}}
\newcolumntype{R}[1]{>{\PreserveBackslash\raggedleft}p{#1}}
\newcolumntype{L}[1]{>{\PreserveBackslash\raggedright}p{#1}}
\newcommand{\RNum}[1]{\uppercase\expandafter{\romannumeral #1\relax}}

\preprint{APS/123-QED}

\title{Valley and spin splittings in a valley-layer-locked monolayer via atomic adsorption}

\author{Xiaohui Wang}
\author{Li Liang}
\author{Xiao Li}%
 \email{lixiao@njnu.edu.cn}
\affiliation{%
Center for Quantum Transport and Thermal Energy Science, School of Physics and Technology, Nanjing Normal University, Nanjing 210023, China\\
}%

\date{\today}

\begin{abstract} 

To date, a number of valley materials have been discovered with the spin-valley or valley-layer couplings. It is highly desirable to realize the interplay of more electronic degrees of freedom in a valley material. Based on the first-principles calculations, we demonstrate the valley and spin degeneracy liftings in the band structure of a TiSiCO monolayer, with the help of the selective expression of the layer pseudospin via atomic adsorption. The introduction of the transition-metal adatoms provides an effective electric field and magnetic proximity effect, giving rise to the valley and spin splittings. These splittings can be further tuned by applying an external electric field. According to the modified band structure, various interlayer excitons with different combinations of spins and electric dipoles are selectively created, under the optical field of appropriate frequencies. The tunable spin and valley splittings in atom adsorbed TiSiCO monolayer offer opportunities for exploring the interactions between spin, valley and layer pseudospin, and designing advanced optoelectronic devices.
\end{abstract}

\maketitle


\section{\label{sec:level1}INTRODUCTION}

Valley, as a novel electronic degree of freedom, has attracted widespread attentions in recent years \cite{schaibley2016valleytronics, vitale2018valleytronics, liu2019valleytronics}. Monolayer transition-metal dichalcogenides are a class of typical valley material, which are characterized by the coupling between the spin and valley due to a large spin-orbit interaction \cite{cao2012valley, xiao2012coupled, macneill2015breaking, qi2015giant, chen2017valley, zhu2011giant, li2020out}. The spin-valley coupling gives rise to many interesting consequences, e.g. spin-valley-contrasting circular dichroism and anomalous Hall effects. Besides transition-metal dichalcogenides, the spin-valley physics are also found in monolayer $\text{MnPX}_3$ (X = S, Se) \cite{li2013coupling}, $\text{VSi}_2\text{N}_4$ \cite{cui2021spin}, $\text{V}_2\text{Se}_2\text{O}$ \cite{ma2021multifunctional} etc. On the other hand, the coupling between valley and layer has been proposed in two-dimensional materials, e.g. monolayer TiSiCO \cite{yu2020valley}, bilayer $\text{MnF}_4$ and bilayer $\text{TaOI}_2$ \cite{xu2021stable}, where two valley states are contributed by different atomic layers. The valley-layer coupling enables the creation of interlayer excitons and the energy-efficient control of the valley polarization via an external electric field.

Given that both the spin-valley and the valley-layer couplings exhibit conceptual importance and application potential, it is highly intriguing that a two-dimensional electronic system can host both spin and layer pseudospin in a valley structure and thus give full play to multiple electronic degrees of freedom. This kind of systems can be realized by constructing a bilayer or multilayer structure composed of the spin-valley-coupled monolayers, where bilayer transition-metal dichalcogenides are typical examples \cite{wu2013electrical, jiang2017zeeman, ciccarino2019carrier}. Besides, as an alternate way of designing the spin-valley-layer related systems, it is likely to introduce the spin degree of freedom into the valley-layer-locked materials. However, for the alternate way, the studies are lacking on not only its material realization but also associated physics. 

In this paper, taking the TiSiCO monolayer as an example of the valley-layer-locked materials, our first-principle calculations demonstrate that the interactions between spin, valley and layer in the monolayer via the atomic adsorption. There are considerable valley splittings and spin splittings in the band structure. These splittings results from the distinct shifts of different band-edge states associated with different atomic layers and different spins, under the action of the transition-metal adatom. To be specific, it is the combined roles of the effective vertical electric field produced by the atomic adsorption and the valley-layer coupling of the band-edge states that lead to the valley splittings. The spin splittings are realized by the magnetic proximity effect from the adatoms, and their magnitudes are larger for the band-edge states composed of the atomic layer closer to adatoms. The valley and spin splittings can be further regulated by an applied electric field. In the presence of the valley and spin degeneracy liftings, non-polarized optical fields with certain frequencies can be used to selectively excite different band-edge states and generate multiple types of interlayer excitons with opposite spins and opposite electric dipoles. The intriguing findings broaden the scope of the research of valleytronics and spintronics materials, and pave the way for designing energy-efficient devices based on the couplings of multiple electronic degrees of freedom.

\section{Methods}
First-principles density-functional-theory calculations \cite{hohenberg1964density, kohn1965self}, as implemented in VASP \cite{kresse1996efficient, kresse1996efficiency}, are performed to relax atomic structures and investigate electronic properties of transition-metal adsorbed TiSiCO monolayer. The projector augmented wave potentials \cite{bliichl1994projector} and the Perdew-Burke-Ernzerhof exchange-correlation functionals \cite{perdew1996generalized} are used with a plane-wave cutoff of 520 eV. A 5$\times$5$\times$1 Monkhorst-Pack k-mesh \cite{monkhorst1976special} is adapted to sample the Brillouin zone of the TiSiCO supercell. A vacuum layer, with a thickness of 17 \AA, is inserted to minimize the interaction between the two-dimensional layer and its periodic images. The convergence thresholds of the total energy and the interatomic force on each atom are set to $\text{10}^{-5}$ eV and 0.01 eV/\AA, respectively. In electronic structure calculations, the HSE06 hybrid functional \cite{heyd2003hybrid} is further employed to more accurately account for the band gap of the semiconductor monolayer, and the vertical electric field is applied by adding a sawtooth potential \cite{makov1995periodic, neugebauer1992adsorbate}.

For each adsorption configuration of transition-metal atom on TiSiCO monolayer, we calculate the adsorption energy, $E_\text{ads}$, by the relation,
\begin{eqnarray}
{E_\text{ads}}={E_\text{tot}}{-E_\text{TSCO}}{-E_\text{TM}},
\end{eqnarray}
where $E_\text{tot}$ is the total energy of relaxed adsorption configuration. $E_\text{TSCO}$ and $E_\text{TM}$ are the energies of a free-standing TiSiCO monolayer and an isolated transition-metal atom, respectively.
To quantify the spin splitting for a given band (the valence or conduction band) at a given valley,  $\Delta^{v/c,\tau}_\text{spin}$, and the valley splitting for a given band, ${\Delta^{v/c}_\text{val}}$, in the band structure, we define the relations
\begin{eqnarray}
{\Delta^{v/c,\tau}_\text{spin}}={E^{v/c,\tau}_\text{$\uparrow$}}{-E^{v/c,\tau}_\text{$\downarrow$}},
\\
{\Delta^{v/c}_\text{val}}={E^{v/c,+1}_\text{val}}{-E^{v/c,-1}_\text{val}},
\end{eqnarray}
where $v$ and $c$ represent the valence and conduction bands, respectively. $\tau=\pm$1 denote the X and X$^{\prime}$ valleys, respectively. $\uparrow$ and $\downarrow$ distinguish the spin-up and spin-down states. $E^{v/c,\tau}_{\uparrow/\downarrow}$ is the energy of the corresponding electronic state. $E^{v/c,\tau}_\text{val}$, chosen from $E^{v/c,\tau}_{\uparrow}$ and $E^{v/c,\tau}_{\downarrow}$, is the one closer to the Fermi level.

\section{Results}

\subsection{Atomic structure of cobalt adsorbed TiSiCO}

We first take the cobalt atom for example to demonstrate the effects of transition-metal adsorption on the atomic and electronic structures of TiSiCO monolayer. The results of other adatoms, iron and nickel, will be given at the end of the section. To study the adsorption system, we construct a 3$\times$3 supercell of the TiSiCO monolayer, as shown in Fig. 1. The monolayer has a square lattice and a space group of P$\overline{4}$m2 \cite{yu2020valley}. It includes five atomic layers with a stacking order of O-Ti-Si/C-Ti-O. The in-plane lattice constant of the TiSiCO supercell is computed to be 8.45\AA.

\begin{figure}[htpb]
\includegraphics[width=85mm]{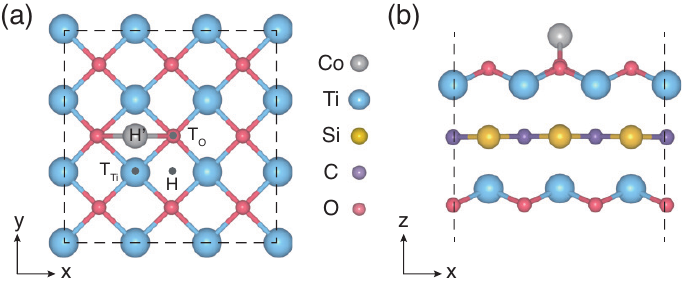}
\caption{\label{fig:graph} Atomic structures of the cobalt adsorbed TiSiCO monolayer. (a) Top view and (b) side view of the most stable adsorption configuration. The supercell of the adsorption system is bounded by dashed lines. Different elements are represented by balls with different colors and sizes. The black dots denote possible adsorption sites. In order to clearly demonstrate the adsorption sites, only the upper O and Ti atomic layers of the monolayer are given in (a).}
\end{figure}

On the top surface of the monolayer, we consider four highly-symmetric adsorption sites. They are two top sites and two hollow sites in the square mesh of the upper Ti and O atomic layers [See Fig. 1(a)]. The two top sites, T$_\text{Ti}$ and T$_\text{O}$, are located above the upper Ti and O atoms, respectively, while the in-plane coordinates of two hollow sites, H and H$^{\prime}$, coincide with those of C and Si atoms in the middle atomic layer, respectively. The computed adsorption energies of these adsorption configurations are in the range from $-$0.8 eV to $-$1.8 eV, indicating that these configurations are (meta-) stable. Among these configurations, the H$^{\prime}$ site is the most stable with the lowest adsorption energy, and its top view and side view are given in Figs. 1(a) and (b), respectively. Herein, the in-plane lattice constant of the fully relaxed supercell is changed by only 0.3 \%, compared with the pristine TiSiCO monolayer. The nearest neighbors of the cobalt atom are two oxygen atoms, and the distance between the cobalt and the oxygen atoms is 1.91 \AA, exhibiting a chemical bonding. Owing to the chemical bonding, the two oxygen atoms are shifted upwards by 0.11 \AA, compared with the other oxygen atoms. The related structural parameters are also listed in Table \RNum{1}.

\begin{table}[b]
\caption{\label{tab:table1}
The most stable configuration of the cobalt adsorbed TiSiCO monolayer and corresponding physical parameters. The vertical height from the adatom to the top oxygen layer ($h$), the bond length between the adatom and the nearest-neighbor oxygen atom ($d$), the adsorption energy ($ E_\text{ads} $), the total magnetic moment in a supercell ($ \mu_\text{tot} $), and the magnetic moment of the adatom ($ \mu_\text{TM} $) are listed in the second row, respectively. Valley and spin splittings in the electronic band structure are given in the fourth row, where the splitting energies are in units of meV.}
\begin{ruledtabular}
\begin{tabular}{C{1.2cm}C{1.2cm}C{1.2cm}C{1.2cm}C{1.2cm}C{1.2cm}}
 \makecell*[c]{Site} & \makecell[c]{$ h $\\(\AA)} & \makecell[c]{$ d $\\(\AA)} & \makecell[c]{$ E_\text{ads}$\\(eV)} & \makecell[c]{$ \mu _\text{tot}$\\($\mu_\text{B}$)} & \makecell[c]{$ \mu_\text{TM} $\\($\mu_\text{B}$)}\\
\hline
\specialrule{0em}{2pt}{2pt}
 H$^{\prime}$ & 1.19 & 1.91 & $-$1.72 & 0.95 & 1.21 \\
\specialrule{0em}{1pt}{1pt}
\hline
\hline
\specialrule{0em}{2pt}{2pt}
 $ \Delta^\text{v}_\text{val}$ & $ \Delta^\text{c}_\text{val}$ & $ \Delta^\text{v,$+$}_\text{spin}$ & $ \Delta^\text{c,$+$}_\text{spin}$ & $ \Delta^\text{v,$-$}_\text{spin}$ & $ \Delta^\text{c,$-$}_\text{spin}$ \\
\specialrule{0em}{2pt}{2pt}
\hline
\specialrule{0em}{2pt}{2pt}
$-$190 & 134 & 85 & $-$4 & $-$12 & 148  \\
\end{tabular}
\end{ruledtabular}

\end{table}

\subsection{Electronic properties of cobalt adsorbed TiSiCO}

For the stable adsorption configuration, we further calculated its magnetic and electronic properties. After the cobalt adsorption, the system has a total magnetic moment of 0.95 $\mu_\text{B}$ in a supercell. This is mainly contributed by the cobalt atom, of which the magnetic moment is 1.21 $\mu_\text{B}$. There are also small contributions from the upper O and Ti layers by the magnetic proximity effect.

\begin{figure}[b]
\includegraphics[width=85mm]{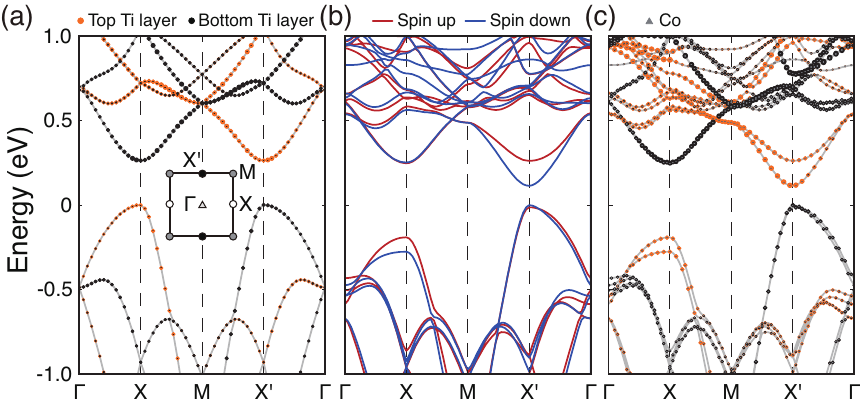}
\caption{\label{fig:graph} Band structures of the TiSiCO monolayer before and after the cobalt adsorption. (a) The atom-projected band structure of the pristine TiSiCO monolayer. (b) Spin-resolved and (c) atom-projected band structures after the cobalt adsorption. The orange and black circles in (a) and (c) are used to indicate the atom-projected weights on the top and bottom Ti layers, respectively, while the newly introduced grey triangles in (c) correspond to the weight on the adsorbed cobalt. The inset of (a) shows the Brillouin zone of the square lattice and its highly-symmetric points. In (b), red and blue lines represent spin-up and spin-down states, respectively. The Fermi level is set to 0 eV.}
\end{figure}

Fig. 2 demonstrates the band structures of the TiSiCO monolayer, before and after the cobalt adsorption. Herein, the spin-orbit coupling is not taken into account in the band structures, since it has no qualitative effect on the results, which will be discussed in Supplemental Material (SM hereafter). As for the pristine non-magnetic TiSiCO monolayer, the bands in Fig. 2(a) are all spin-degenerate. There is a pair of degenerate valleys, which are located at the highly-symmetric points, X and X$^{\prime}$, of the square Brillouin zone [See the inset of Fig. 2(a)]. Owing to the valley degeneracy, the direct band gaps at two valleys are equal, with a magnitude of 0.26 eV. According to the atomic projection, the band-edge states at both valleys are mainly contributed by the Ti atoms. The valence band states at X valley and the conduction band states at X$^{\prime}$ valley are both contributed by the top Ti layer, while the valence band states at X$^{\prime}$ valley and the conduction band states at X valley are both composed of the bottom Ti layer. That is, the degenerate band-edge states correspond to two Ti layers, respectively, demonstrating a valley-layer coupling in 3$\times$3 TiSiCO supercell, which is consistent with previous results of the unit cell of TiSiCO \cite{yu2020valley}. 

After the cobalt adsorption, the valley structure is well preserved in the energy range from $-$0.4 eV to $+$0.4 eV, without introducing any additional bands, as shown in Fig. 2(b). However, in the valley structure of the cobalt adsorbed TiSiCO monolayer, the valley degeneracy is lifted. The induced valley splittings have magnitudes of $-$190 meV and $+$134 meV for the valence and conduction bands, respectively, where the positive/negative sign denotes a higher energy level at X/X$^{\prime}$ valley. As a result, the global valence band maximum (VBM) and conduction band minimum (CBM) are both located at X$^{\prime}$ valley, with a direct band gap of 0.11 eV. Moreover, it is seen that the spin degeneracy of each band is also lifted, by the spin projection. Among all the low-energy bands, the spin splittings are larger for the valence band at X valley and the conduction band at X$^{\prime}$ valley, with magnitudes of $+$85 meV and $+$148 meV, respectively. Here, the positive sign means the spin-up state has higher energy compared with its spin-down counterpart. The spin splittings of the other valence and conduction band states are also given in Table \RNum{1}. Because of these spin splittings, the local VBM and CBM at X valley are both spin-up, and those at X$^{\prime}$ valley are both spin-down, exhibiting a spin-valley coupling arising from the magnetic adatoms. Besides, if we only focus on the global VBM and CBM, the same spin at the two band edges indicates that the cobalt adsorbed TiSiCO monolayer is a half-semiconductor \cite{rudberg2007nonlocal, yoo2010spin}.

To further elucidate the splittings of the valley and spin degeneracies in the band structure of the cobalt adsorbed TiSiCO monolayer, we analyze the atom projections of the band structure and the Bader charge transfer. Similar to the pristine monolayer, the atom-projected band edge states of the adsorption system in Fig. 2(c) also exhibit the valley-layer coupling, i.e. two valleys are mainly contributed by different Ti atomic layers, for both the valence and conduction bands. Besides, the band-edge states with large spin splittings have also small orbital contributions from the adatom. Electronic states mainly composed of the cobalt atom are located in the energy range below $-$1.
2eV. On the other hand, the Bader charge analysis demonstrates that 0.44 electron is transferred from the cobalt atom to the TiSiCO monolayer. The charge transfer induces a built-in electric field pointing from the positively charged adatom to the negatively charged monolayer. Under the action of the electric field, the electrons in the top Ti layer feel a smaller electric potential energy than those in the bottom Ti layer. Considering the additional electric potential energy, the VBM at X valley, contributed by the top Ti layer, apparently becomes lower than the one at X$^{\prime}$ valley, contributed by the bottom Ti layer. In a similar manner, the CBM at X$^{\prime}$ valley has smaller electronic energy than the one at X valley. The above relative band shifts lead to valley splittings. Meanwhile, the magnetic moment from the cobalt atom results in the spin splitting of the band-edge states. Since the top Ti layer is closer to the cobalt atom than the bottom Ti layer, the cobalt orbitals are more involved in the band-edge states contributed by the top Ti layer and bring these states larger spin splittings. In a word, the magnetic adatom introduces both the effective electric field and the magnetic proximity effect, giving rise to the valley and spin splittings in the band structure.
\subsection{The effect from the external electric field}

Given that the cobalt adsorption plays the roles of effective fields, it is possible for an external vertical electric field to further tune the valley and spin splittings. We thus study the effect of the vertical electric field on the band-edge states. Fig. 3(a) shows the evolutions of the valley splittings as functions of the electric field. Herein, an electric field along the $+$z axis is set to positive. It is found that when sweeping the strength of the electric field from 0 V/\AA\ to $+$0.2 V/\AA, the negative splitting of the valence band constantly increases, while the positive splitting of the conduction band decreases. That is, the absolute values of the valley splittings for both bands are continually reduced. In contrast, the absolute values keep on increasing when the electric field is swept from 0 V/\AA\ to $-$0.2 V/\AA. To be specific, an electric field of $+$0.2 V/\AA ($-$0.2 V/\AA) changes the valley splittings by $+$66 meV ($-$34 meV) and $-$84 meV ($+$46 meV) for the valence and conduction bands, respectively, compared with the case without the electric field. Therefore, the electric field is indeed an effective way to further modulate the valley splitting in the cobalt adsorbed TiSiCO monolayer.

\begin{figure}[htpb]
\includegraphics[width=85mm]{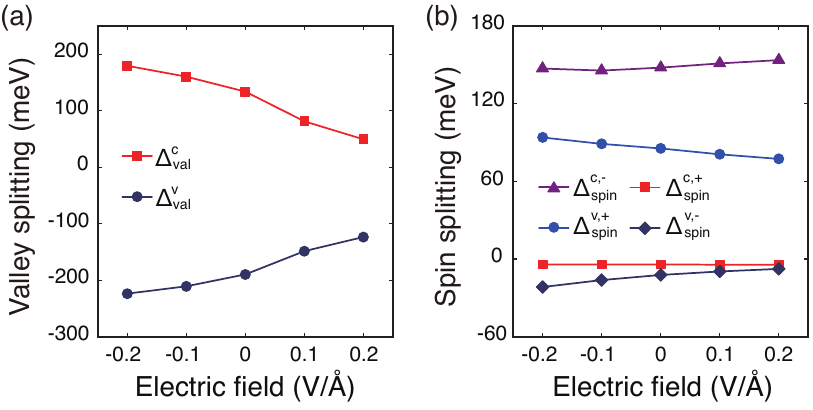}
\caption{\label{fig:graph} The evolutions of the valley and spin splittings as functions of an applied vertical electric field. (a) The valley splittings of the valence and conduction bands. (b) The spin splittings of both bands at both valleys.}
\end{figure}

The change of the valley splittings is attributed to the combined roles of the built-in electric field from the charge transfer and the applied electric field. When the upward external electric field is opposite to the built-in one, the cancellation of the roles from two electric fields leads to decreases in the sizes of the valley splittings for both the valence and conduction bands. When the downward external electric field has the same direction with the built-in one, the external electric field helps strengthen the magnitudes of the valley splittings for both bands.

On the other hand, Fig. 3(b) shows the changes of the spin splittings of the valence and conduction bands at two valleys with the applied electric field. Different band-edge states exhibit distinct change trends of the spin splittings. Besides, for all band-edge states, the changes of the spin splittings have values of less than 10 meV within the strength range of the electric field considered here, and correspondingly the changes are smaller compared with the case of the valley splittings.

\subsection{Valley optical transition}

Given the valley and spin splittings in the electronic band structure of the cobalt adsorbed TiSiCO monolayer, subsequent optical consequences become compelling. The above band degeneracy liftings result in the energy splittings of optical transitions between band-edge states. Assuming the spin flip is forbidden, there are four possible direct optical transitions, which, together with corresponding transition energies, are shown in Fig. 4(a). These direct optical transitions can be selectively excited by tuning optical frequency, regardless of the optical polarization. That is, the linearly polarized, circularly polarized, and unpolarized lights can all be adopted as long as the optical frequency is appropriate. In contrast, for the pristine TiSiCO monolayer, only linearly polarized light can be used to selectively excite two degenerate valleys.

\begin{figure}[htpb]
\includegraphics[width=85mm]{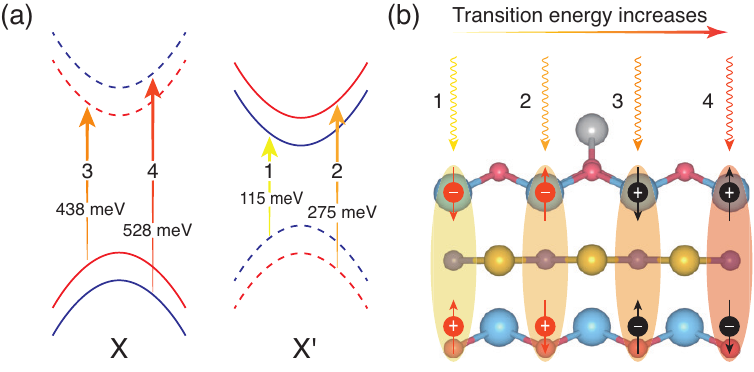}
\caption{\label{fig:graph} Valley optical transitions and associated interlayer excitons in the cobalt adsorbed TiSiCO monolayer. (a) Four possible direct optical transitions without considering the spin flip. The solid and dashed lines stand for the band-edge states contributed by the top and bottom Ti layers, respectively. Spin-up and spin-down states are denoted in red and blue, respectively. The numbers 1-4, along with corresponding optical transition energies nearby, distinguish different optical transitions between these band-edge states. (b) Interlayer excitons generated by the optical transitions in (a). Electrons and holes at X valley are represented by  orange circles with $-$ and $+$ symbols, respectively. The counterparts in X$^{\prime}$ valley are represented by black circles. The up and down arrows on the circles stand for the up and down spins, respectively. These electrons and holes are localized on the top or bottom Ti layer.}
\end{figure}

As a result of the optical transitions, four kinds of excitons can be selectively created with distinct combinations of the spins and electric polarizations in the cobalt adsorbed TiSiCO monolayer, as shown in Fig. 4(b). Each kind of exciton is formed by a pair of electron and hole with opposite spins. Further taking into account the spatial distributions of band-edge states, the excitons are interlayer ones, with the electrons and holes lying in different Ti atomic layers, giving rise to an electric polarization along the $+z$ or $-z$ direction. Taking the spin-down valence and conduction bands at X$^{\prime}$ valley for an example, spin-down electrons and spin-up holes are excited under an optical field of 115 meV to produce the interlayer excitons labeled by 1 in Fig. 4(b). This kind of excitons, with holes/electrons being distributed at the top/bottom Ti layer, exhibits an electric dipole pointing upwards. Similarly, the other interlayer excitons labeled by 2-4, with opposite spin or/and opposite electric polarization, can also be readily available by applying optical fields of appropriate energies.

\begin{table}[htpb]
\caption{\label{tab:table6}
Physical properties of iron and nickel adsorbed TiSiCO monolayers. The properties include the most stable adsorption site, structural parameters, adsorption energy, magnetic moments, valley splittings, and spin splittings. The meaning of every item can be found in the caption of Table \RNum{1}. The splitting energies are given in units of meV.
}
\begin{ruledtabular}
\centering
\begin{tabular}{ m{1cm}<{\centering} m{1cm}<{\centering} m{1cm}<{\centering} m{1cm}<{\centering} m{1cm}<{\centering} m{1cm}<{\centering} m{1cm}<{\centering}}
  & \makecell*[c]{Site} & \makecell[c]{$ h $\\(\AA)} & \makecell[c]{$ d $\\(\AA)} & \makecell[c]{$ E_\text{ads}$\\(eV)} &\makecell[c]{$ \mu _\text{tot}$\\($\mu_\text{B}$)} &\makecell[c]{$ \mu_\text{TM}$\\($\mu_\text{B}$)}\\
\hline
\specialrule{0em}{2pt}{2pt}
Fe & H & 1.51 & 2.08 & $-$1.50 & 3.42 & 3.35 \\
Ni & H$^{\prime}$ & 1.25 & 1.94 & $-$1.73 & 0.00 & 0.00 \\
\specialrule{0em}{1pt}{1pt}
\hline
\hline
\specialrule{0em}{2pt}{2pt}
 & $ \Delta^\text{v}_\text{val}$ & $ \Delta^\text{c}_\text{val}$ & $ \Delta^\text{v,$+$}_\text{spin}$ & $ \Delta^\text{c,$+$}_\text{spin}$ &  $ \Delta^\text{v,$-$}_\text{spin}$ & $ \Delta^\text{c,$-$}_\text{spin}$ \\
\specialrule{0em}{2pt}{2pt}
\hline
\specialrule{0em}{2pt}{2pt}
Fe & $-$66 & $-$6 & $-$126 & 8 & 0 & 27  \\
Ni & $-$183 & 84 & 0 & 0 & 0 & 0  \\
\end{tabular}
\end{ruledtabular}

\end{table}

\subsection{Electronic properties of iron and nickel adsorptions}
In addition to the cobalt adsorption, we also calculated the iron and nickel adsorbed TiSiCO monolayers. The structural parameters and magnetic properties of these adsorptions are listed in Table \RNum{2}. The most stable adsorption sites are the H site for the iron atom and the H$^{\prime}$ site for the nickel atom. The chemical bonds are also formed between the adsorbed atom and the nearest-neighboring oxygen atoms, which are characterized by a large adsorption energy of $-$1.50 eV ($-$1.73 eV) and short bond lengths of 2.08\AA\ (1.94 \AA) for the iron (nickel) adsorption. Owing to the chemical bonding, the nearest-neighboring oxygen atoms are shifted upwards by 0.09\AA\ for both the iron and nickel adsorptions, compared to the other oxygen atoms. Moreover, while the magnetic moment of the adsorbed iron is 3.42 $\mu_\text{B}$, the nickel atom has not produced any magnetic moment in the adsorption system.

The iron and nickel adsorptions also lead to the modification of the band structures. The corresponding band structures are given in S.I., and the values of the valley and spin splittings are also listed in Table \RNum{2}. Similar to the cobalt adsorption, the iron adsorption produces both valley and spin splittings. The valley splitting of the valence band is considerable, with a value of $-$66 meV. Larger spin splittings are also realized in the band-edge states contributed by the top Ti layer. On the other hand, the nickel adsorption exhibits sizable valley splittings, with values of $-$183 meV and 84 meV for the valence and conduction bands, respectively. However, there is no spin splitting due to the vanishing magnetic moment of the adsorbed nickel.

\section{Conclusion}
In summary, we investigated the effects of transition-metal atom adsorption on the electronic structures of the TiSiCO monolayer. The transition-metal atoms include iron, cobalt, and nickel. Their adsorptions lead to valley splittings, due to the proximity-induced effective electric field and the valley-layer coupling of the TiSiCO monolayer. In particular, the iron and cobalt adsorptions also induce magnetic moments and associated effective Zeeman field, giving rise to spin splittings of the band-edge states. Moreover, it is found that the valley and spin splittings can be further tuned by an applied vertical electric field. The modified electronic structures after the atomic adsorptions enable the selective excitations of interlayer excitons with various combinations of spins and electric polarizations, by tuning the frequency of the optical field. These findings demonstrate an intriguing interplay of the spin, valley, and layer degrees of freedom in the transition-metal atom adsorbed TiSiCO monolayer, and provide an avenue for exploring advanced valleytronics and spintronics devices. 

\bigskip
\begin{acknowledgments}
We acknowledge
financial supports from the National Natural Science Foundation of China Grant 11904173
and the Jiangsu Specially-Appointed Professor Program.
\end{acknowledgments}

\nocite{*}

\providecommand{\noopsort}[1]{}\providecommand{\singleletter}[1]{#1}%

\end{document}